\documentclass{aa}  
\usepackage{graphicx}
\usepackage{txfonts}
\usepackage{xcolor}
\usepackage{hyperref}
\hypersetup{colorlinks, linkcolor=blue, citecolor=blue, urlcolor=blue}

\newcommand{\muvec}{\mbox{\boldmath $\mu$}}
\newcommand{\xivec}{\mbox{\boldmath $\xi$}}

\newcommand{\te}{t_{\rm E}}
\newcommand{\thetae}{\theta_{\rm E}}

\newcommand{\pie}{\pi_{\rm E}}

\newcommand{\dl}{D_{\rm L}}

\def\e{{\rm E}}

\definecolor{brown}{rgb}{0.59, 0.29, 0.0}
\definecolor{darkgreen}{rgb}{0.0, 0.42, 0.24}
\definecolor{darkblue}{rgb}{0.01, 0.31, 0.59}
\definecolor{darkblue}{rgb}{0.0, 0.25, 0.42}
\definecolor{blue}{rgb}{0.0,0.0,1.0}
\definecolor{green}{rgb}{0.0,1.0,0.0}

\begin{document}

\title{KMT-2017-BLG-0673Lb and KMT-2019-BLG-0414Lb: Two microlensing planets detected in peripheral fields of KMTNet survey}
\titlerunning{Two KMTNet microlensing planets detected in peripheral fields}

\author{
     Cheongho Han\inst{\ref{01}} 
\and Chung-Uk~Lee\inst{\ref{02}} 
\and Andrew Gould\inst{\ref{03},\ref{04}} 
\and Youn~Kil~Jung\inst{\ref{02}} 
\and Michael~D.~Albrow\inst{\ref{05}} 
\and Sun-Ju~Chung\inst{\ref{02}} 
\and Kyu-Ha~Hwang\inst{\ref{02}} 
\and Doeon~Kim\inst{\ref{01}} 
\and Yoon-Hyun~Ryu\inst{\ref{02}} 
\and In-Gu~Shin\inst{\ref{06}} 
\and Yossi~Shvartzvald,\inst{\ref{07}} 
\and Hongjing Yang\inst{\ref{08}} 
\and Jennifer~C.~Yee\inst{\ref{06}} 
\and Weicheng~Zang\inst{\ref{08}} 
\and Sang-Mok~Cha\inst{\ref{02},\ref{09}} 
\and Dong-Jin~Kim\inst{\ref{02}} 
\and Seung-Lee~Kim\inst{\ref{02}} 
\and Dong-Joo~Lee\inst{\ref{02}}
\and Yongseok~Lee\inst{\ref{02}} 
\and Byeong-Gon~Park\inst{\ref{02}} 
\and Richard~W.~Pogge\inst{\ref{04}}
\\
(The KMTNet Collaboration) \\
}

\institute{
       Department of Physics, Chungbuk National University, Cheongju 28644, Republic of Korea  \\ \email{cheongho@astroph.chungbuk.ac.kr}   \label{01} 
\and   Korea Astronomy and Space Science Institute, Daejon 34055,Republic of Korea                                                          \label{02} 
\and   Max Planck Institute for Astronomy, K\"onigstuhl 17, D-69117 Heidelberg, Germany                                                     \label{03} 
\and   Department of Astronomy, The Ohio State University, 140 W.  18th Ave., Columbus, OH 43210, USA                                       \label{04} 
\and   University of Canterbury, Department of Physics and Astronomy, Private Bag 4800, Christchurch 8020, New Zealand                      \label{05} 
\and   Center for Astrophysics~|~Harvard \& Smithsonian, 60 Garden St., Cambridge, MA 02138, USA                                            \label{06}
\and   Department of Particle Physics and Astrophysics, Weizmann Institute of Science, Rehovot 76100, Israel                                \label{07} 
\and   Department of Astronomy, Tsinghua University, Beijing 100084, China                                                                  \label{08} 
\and   School of Space Research, Kyung Hee University, Yongin, Kyeonggi 17104, Republic of Korea                                            \label{09}  
}
\date{Received ; accepted}

\abstract
{}
{
We investigate the microlensing data collected during the 2017--2019 seasons in the peripheral 
Galactic bulge fields with the aim of finding planetary signals in microlensing light curves 
observed with relatively sparse coverage.
}
{
We first sort out lensing events with weak short-term anomalies in the lensing light curves
from the visual inspection of all non-prime-field events, and then test various interpretations 
of the anomalies. From this procedure, we find two previously unidentified candidate planetary 
lensing events KMT-2017-BLG-0673 and KMT-2019-BLG-0414.  It is found that the planetary signal of 
KMT-2017-BLG-0673 was produced by the source crossing over a planet-induced caustic, but it was 
previously missed because of the sparse coverage of the signal.  On the other hand, the possibly 
planetary signal of KMT-2019-BLG-0414 was generated without caustic crossing, and it was previously 
missed due to the weakness of the signal.  We identify a unique planetary solution for 
KMT-2017-BLG-0673. However, for KMT-2019-BLG-0414, we identify two pairs of planetary solutions, 
for each of which there are two solutions caused by the close-wide degeneracy, and a slightly 
less favored binary-source solution, in which a single lens mass gravitationally magnified a 
rapidly orbiting binary source with a faint companion (xallarap).
}
{
From Bayesian analyses, it is estimated that the planet KMT-2017-BLG-0673Lb has a mass of 
$3.7^{+2.2}_{-2.1}~M_{\rm J}$, and it is orbiting a late K-type host star with a mass of 
$0.63^{+0.37}_{-0.35}~M_\odot$.  Under the planetary interpretation of KMT-2010-BLG-0414L,
a star with a mass of $0.74^{+0.43}_{-0.38}~M_\odot$ hosts a planet with a mass of 
$\sim 3.2$--3.6~$M_{\rm J}$ depending on the solution.  We discuss the possible resolution 
of the planet-xallarap degeneracy of KMT-2019-BLG-0414 by future adaptive-optics observations 
on 30~m class telescopes.  The detections of the planets indicate the need for thorough 
investigations of non-prime-field lensing events for the complete census of microlensing 
planet samples.
}
{}


\maketitle

\section{Introduction}\label{sec:one}

One of the most important scientific features of planetary microlensing is that it is 
sensitive to cold planets lying at around or beyond the snow line. Detecting these 
planets is important because, according to the core-accretion theory of the planet 
formation \citep{Laughlin2004}, there are abundant solid grains at around the snow line 
to be accreted into planetesimals that will eventually evolve into gas giants  by accreting 
gas residing in the surrounding disk. It is, in general, difficult to detect these cold 
planets using other major planet detection methods such as the radial-velocity and 
transit methods because of the long orbital periods of the planets. Hence, the microlensing 
method is essential for the demographic studies that encompass planets distributed throughout 
a wide region of planetary systems \citep{Gaudi2012}.

For the complete demographics of planets, it is important to accurately estimate the 
planet detection efficiency. The efficiency of microlensing planets is calculated as 
the ratio of the number of detected planets to the total number of lensing events. If 
a fraction of planets are missed despite their signals being above a detection threshold, 
then their frequency would be underestimated, and this would lead to an erroneous
evaluation of planet properties.

Strong planetary signals appearing in lensing light curves covered with high cadences
and good photometric precision, for example, OGLE-2018-BLG-1269 \citep{Jung2020a} and  
KMT-2019-BLG-0842 \citep{Jung2020b}, can be readily identified. However, identifying planetary 
signals for a subset of lensing events is difficult caused by various reasons, such as very 
short durations of signals, low photometric precision due to the faintness of the source or 
the occurrence of planetary signals during low lensing magnifications, weakness of signals 
due to their non-caustic-crossing nature, and relatively sparse coverage of signals for 
events detected in low-cadence fields and (or) in the wings of the observing season.  In 
some cases of events produced by high-mass planets lying in the vicinity of the Einstein 
ring, the planet-induced caustics have a resonant form.  In this case, the planetary signals 
can significantly deviate from the typical form of a short-term anomaly \citep{GouldLoeb1992}, 
and this also makes it difficult to immediately identify the planetary signals. Therefore, 
finding unidentified planetary signals in the data of lensing surveys is important for the 
accurate estimation of planet frequency.

Searches for unidentified microlensing planets in the data collected by the Korea Microlensing
Telescope Network \citep[KMTNet:][]{Kim2016} survey have been carried out in two major channels. 
The first channel is a systematic investigation of the residuals in lensing light curves from 
the single-lens single-source (1L1S) fits. This approach has been applied to the KMTNet 
prime-field data obtained in the 2018 and 2019 seasons, leading to the discoveries of dozens 
of planets: 1 planet (OGLE-2019-BLG-1053Lb) published in \citet{Zang2021}, 6 planets 
(OGLE-2018-BLG-0977Lb, OGLE-2018-BLG-0506Lb, OGLE-2018-BLG-0516Lb, OGLE-2019-BLG-1492Lb,
 KMT-2019-BLG-0253Lb, KMT-2019-BLG-0953Lb) published in \citet{Hwang2022}, 1 planet 
(OGLE-2018-BLG-0383Lb) published in \citet{Wang2022}, 3 planets (KMT-2019-BLG-1042Lb, 
KMT-2019-BLG-1552Lb, KMT-2019-BLG-2974Lb) published in \citet{Zang2022}, and 8 planets 
(OGLE-2018-BLG-1126Lb, KMT-2018-BLG-2004Lb, OGLE-2018-BLG-1647Lb, OGLE-2018-BLG-1367Lb, 
OGLE-2018-BLG-1544Lb, OGLE-2018-BLG-0932Lb, OGLE-2018-BLG-1212Lb, and KMT-2018-BLG-2718Lb)  
published in \citet{Gould2022}.  This approach  is being expanded to the data obtained from 
the other fields and to those acquired in other seasons, and 6 planets (KMT-2018-BLG-0030Lb, 
KMT-2018-BLG-0087Lb, KMT-2018-BLG-0247Lb, OGLE-2018-BLG-0298Lb, KMT-2018-BLG-2602Lb, 
OGLE-2018-BLG-1119Lb)  were recently reported from the investigation of the 2018 
sub-prime-field data \citep{Jung2022}.

The second approach for exhuming missing planets is visually inspecting planetary signals. 
\citet{Han2020} found 4 microlensing planets (KMT-2016-BLG-2364Lb, KMT-2016-BLG-2397Lb,
OGLE-2017-BLG-0604Lb, and OGLE-2017-BLG-1375Lb) from visually inspecting faint-source events
found during the 2016 and 2017 seasons. \citet{Han2021a} inspected anomalies with no
caustic-crossing features and identified 3 planets (KMT-2018-BLG-1976Lb, KMT-2018-BLG-1996,
and OGLE-2019-BLG-0954). \citet{Han2021b} additionally identified 3 planets (KMT-2017-BLG-2509Lb, 
OGLE-2017-BLG-1099Lb, and OGLE-2019-BLG-0299Lb) detected via resonant-caustic channels. 
\citet{Han2022} reexamined high-magnification microlensing events in the 2018 season data 
and found 1 planet (KMT-2018-BLG-1988Lb) with a very short-duration planetary signal.

This independent visual search for planetary signals complements the AnomalyFinder search 
and provides a verification sample for evaluating the effectiveness of the AnomalyFinder 
algorithm. The visual search requires modeling a large number of anomalous events to isolate 
the ones of interest. The output of the AnomalyFinder is then compared to this modeling to 
verify that the anomalies have been correctly classified and vet for ambiguous events, such 
as those that suffer from various types of degeneracy. In addition, planets discovered by-eye 
but not recovered by the AnomalyFinder search provide important insight into false negatives 
in that search.

In this work, we report the discovery of a planet (KMT-2017-BLG-0673Lb) and a planet candidate
(KMT-2019-BLG-0414Lb).  These were both identified from the visual inspection of the KMTNet 
data obtained in the peripheral fields, toward which the observational cadences are 
substantially lower than that of the prime fields, and thus planetary signals were sparsely 
covered. KMT-2019-BLG-0414Lb was also identified by the AnomalyFinder search.  However, 
while the anomaly in KMT-2017-BLG-0673 was found by the algorithm, it was rejected by the 
operator because the peak data appeared to be ``noisy'', which, in fact, was due to the 
planetary signal. This event demonstrates the importance and complementarity of such by-eye 
searches.

To present the analyses of the planetary events, we organize the paper as follows. In 
Section~\ref{sec:two}, we describe the observations of the planetary lensing events and 
the procedure of data reduction.  In Section~\ref{sec:three}, we describe the detailed 
features of the observed anomalies, and present analyses of the lensing events conducted 
to explain the anomalies.  In particular, we show that there is an alternate, 
non-planetary, solution for KMT-2019-BLG-0414, in which the anomaly is explained by an 
orbiting companion of the source (xallarap) and which is disfavored by only $\Delta\chi^2=4.2$.  
We explain the procedure of modeling and present the lensing parameters constrained by the 
modeling. In Section~\ref{sec:four}, we specify the source stars of the events and estimate 
angular Einstein radii. In Section~\ref{sec:five}, we estimate the physical lens parameters 
by conducting Bayesian analyses of the events.  In Section~\ref{sec:six}, we discuss how 
future adaptive optics (AO) observations on 30~m class telescopes can resolve the 
planet-xallarap degeneracy for KMT-2019-BLG-0414.  We summarize the results found from 
the analyses in Section~\ref{sec:seven}.

\begin{figure}[t]
\includegraphics[width=\columnwidth]{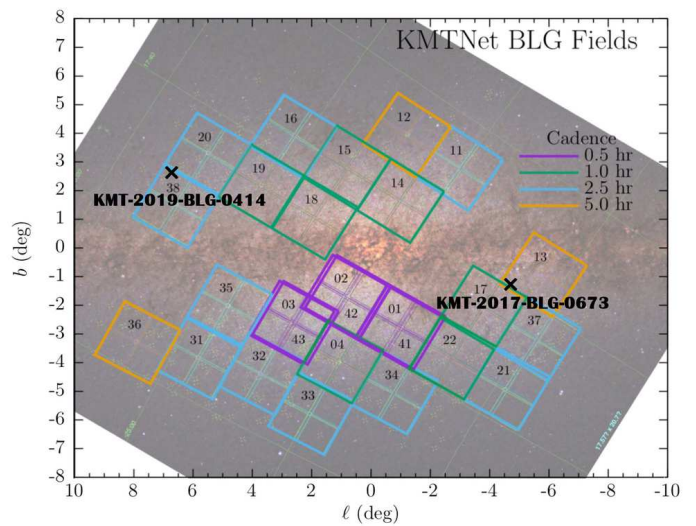}
\caption{
KMTNet fields and cadences.  The locations of the events KMT-2017-BLG-0673 and 
KMT-2019-BLG-0414 are indicated by cross marks. 
}
\label{fig:one}
\end{figure}

\section{Observations}\label{sec:two}

Observations of the KMTNet survey are being carried out in different cadences depending on 
the fields. Six prime fields (KMT01, KMT02, KMT03, KMT41, KMT42, and KMT43) are visited with 
a 0.5~hr cadence. Each pair of the fields KMT01-KMT41, KMT02-KMT42, KMT03-KMT43 overlaps but 
are shifted to cover the gaps between chips in one of the two fields (about 15\% of the total 
region) and thus most region of the prime fields are covered with a 0.25~hr cadence. 
For the other fields, the cadences are lower than that of the prime fields with 1.0~hr cadence 
for 7 fields (KMT04, KMT14, KMT15, KMT17, KMT18, KMT19, and KMT22), 2.5~hr cadence for 11 fields 
(KMT11, KMT16, KMT20, KMT21, KMT31, KMT32, KMT33, KMT34, KMT35, KMT37, and KMT38), and 5~hr 
cadence for 3 fields (KMT12, KMT13, and KMT36).  Figure~\ref{fig:one} shows the fields covered 
by the KMTNet survey and the observational cadences of the individual fields.

The two planetary events KMT-2017-BLG-0673 and KMT-2019-BLG-0414 were found in the KMT13
and KMT38 fields, toward which observations were conducted with 5~hr and 2.5~hr cadences,
respectively. The equatorial coordinates of the individual events are (RA, DEC)$=$(17:39:01.69, 
-33:48:28.91), which correspond to the Galactic coordinates $(l, b)=(-4^\circ\hskip-2pt .882, 
-1^\circ\hskip-2pt.373)$, for KMT-2017-BLG-0673, and (RA, DEC)$=$(17:55:24.70, -21:53:43.94), 
which correspond to $(l, b)=(7^\circ\hskip-2pt .179, 1^\circ\hskip-2pt .709)$, for 
KMT-2019-BLG-0414.  In Figure~\ref{fig:one}, we mark the positions of the two lensing events.

Both events were found solely by the KMTNet survey, and there are no data from the other
currently working surveys of the Optical Gravitational Lensing Experiment 
\citep[OGLE:][]{Udalski2015} and the Microlensing Observations in Astrophysics survey 
\citep[MOA:][]{Bond2001}.  The event KMT-2017-BLG-0673 occurred before the development
of the KMTNet AlertFinder algorithm \citep{Kim2018b}, which began operation in the 2018 season, 
and thus it was identified from the post-season investigation of the data using the KMTNet 
EventFinder algorithm \citep{Kim2018a}.  On the other hand, the event KMT-2019-BLG-0414 
was found using the AlertFinder system in the early stage of the event on 2019 April 16 
(HJD$^\prime \sim 8589.5$), when the apparent source flux was brighter than the baseline 
by $\sim 0.15$~mag.

Observations of both events were done using the three KMTNet telescopes. The individual KMTNet
telescopes are located in three sites of the Southern Hemisphere: Cerro Tololo Interamerican
Observatory in Chile (KMTC), the South African Astronomical Observatory in South Africa (KMTS)
and the Siding Spring Observatory in Australia (KMTA). Each of these telescopes has a 1.6~m
aperture and is equipped with a camera yielding 4~deg$^2$ field of view. Images from the survey
were obtained primarily in the $I$ band, and about 9\% of images were acquired in the $V$
band for the source color measurement.

Reductions of images and photometry of the events were carried out utilizing the automatized
pipeline of the KMTNet survey developed by \citet{Albrow2009}. For a subset of the KMTC data
sets, additional photometry were conducted using pyDIA code \citep{Albrow2017} to measure the 
source colors of the events. The detailed procedure of the source color measurement will be 
discussed in Sect.~\ref{sec:three}. For the data used in the analysis, we readjusted the error 
bars estimated from the photometry pipeline using the method of \citet{Yee2012} so that the 
data are consistent with their scatter and $\chi^2$ per degree of freedom for each data set 
is unity.

\section{Light curve analyses}\label{sec:three}

\begin{figure}[t]
\includegraphics[width=\columnwidth]{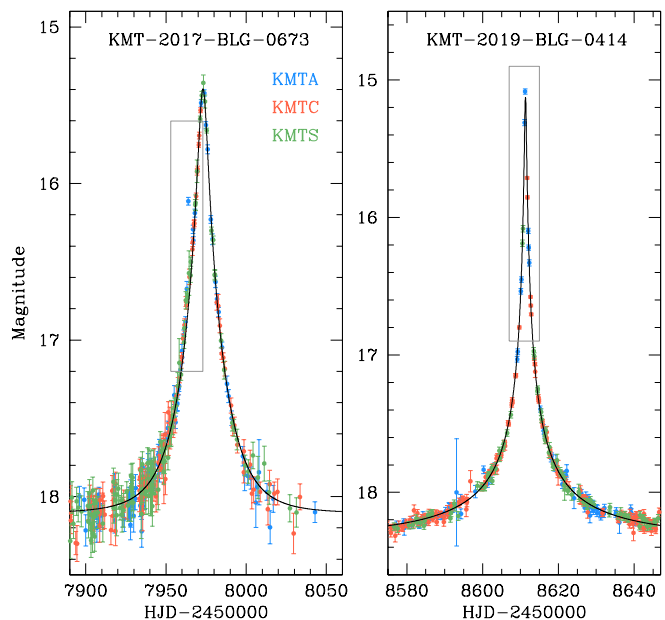}
\caption{
Microlensing light curves of KMT-2017-BLG-0673 and KMT-2019-BLG-0414. The curve drawn over 
the data points of each light curve is a 1L1S model, and the box indicates the region of the 
anomaly.  The enlarged views of the anomaly regions are shown in Fig.~\ref{fig:three} for 
KMT-2017-BLG-0673 and in Fig.~\ref{fig:four} for KMT-2019-BLG-0414.  Colors of data points 
are set to match those of the telescopes marked in the legend. 
}
\label{fig:two}
\end{figure}

\begin{figure}[t]
\includegraphics[width=\columnwidth]{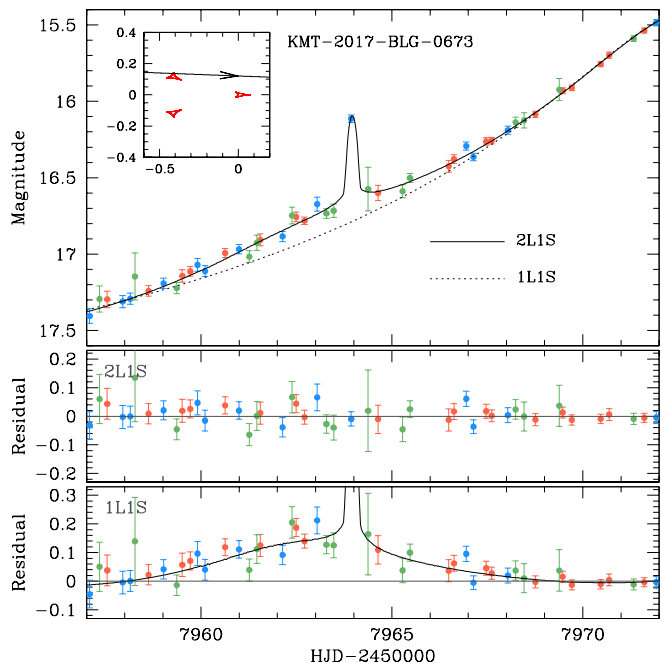}
\caption{
Enlargement of the anomaly region in the lensing light curve of KMT-2017-BLG-0673.  The 
solid and dotted curves drawn over the data points are the 2L1S and 1L1S models, respectively.  
The residuals from the individual models are shown in the lower two panels. The inset in the 
top panel shows the lens system configuration, in which the source trajectory (line with an 
arrow) with respect to the caustic (red figures) is shown.  The curve drawn in the bottom 
panel is the difference between the 2L1S and 1L1S models. 
}
\label{fig:three}
\end{figure}

By inspecting the data of the lensing events detected in the peripheral KMTNet fields covered 
with observational cadences $\geq 1 $~hours during the 2017--2019 seasons, we found that 
KMT-2017-BLG-0673 and KMT-2019-BLG-0414 exhibit subtle deviations from the 1L1S models. 
In Figure~\ref{fig:two}, we present the light curves of the two events.  From a brief glimpse, 
the light curves of both events appear to be well described by 1L1S models, which are drawn 
over the data points. However, a thorough inspection reveal that there exist weak short-term 
anomalies. We mark the regions of the anomalies for the individual events with grey boxes 
drawn over the light curves, and the enlarged views of the anomaly regions are shown in 
Figure~\ref{fig:three} for KMT-2017-BLG-0673 and in Figure~\ref{fig:four} for KMT-2019-BLG-0414.

In the following subsections, we present analyses of both events conducted to explain the observed 
anomalies in the lensing light curves. As will be discussed, the anomalies are well described by a 
binary-lens (2L1S) model, in which the mass ratio between the lens components ($M_1$ and $M_2$) is 
very small.

The 2L1S modeling procedure commonly applied to both events are as follows. In the modeling of 
each event, we  search for a lensing solution, which represents a set of lensing parameters describing 
the observed lensing light curve.  Under the approximation that the relative lens-source motion 
is rectilinear, a 2L1S event light curve is described by 7 basic parameters. The first three of 
these parameters $(t_0, u_0, \te)$ characterize the encounter between the lens and source, and 
the individual parameters denote the time of the closest lens-source approach, the separation 
at that time (impact parameter) normalized to the angular Einstein radius $\thetae$, and the 
event time scale, respectively. The event time scale is defined as the time for the source to 
cross the angular Einstein radius, that is, $\te = \thetae/\mu$, where $\mu$ 
represents the relative lens-source proper motion. The next three parameters $(s, q, \alpha)$ 
define the binary lens, and the first two parameters represent the projected separation (scaled to 
$\thetae$) and mass ratio between $M_1$ and $M_2$, respectively, and the third parameter indicates 
the source trajectory angle defined as the angle between the relative source motion and
the binary axis of the lens. The last parameter $\rho$ (normalized source radius), which is 
defined as the ratio of the angular source radius $\theta_*$ to $\thetae$, characterizes the 
deformation of a lensing light curve by finite-source effects, which occur during the crossing of 
a source over the caustic formed by a binary lens.

The searches for the the best-fit lensing parameters were carried out in two steps. In the first
step, we divided the lensing parameters into two groups, and the binary parameters $(s, q)$ in 
the first group were searched for via a grid approach with multiple initial values of $\alpha$ 
evenly divided in the [0 -- 2$\pi$] range, and the other parameters were found via a downhill 
approach using a Markov Chain Monte Carlo (MCMC) algorithm. We then construct $\chi^2$ maps on 
the $\log s$--$\log q$--$\alpha$ planes, and identify local solutions on the $\chi^2$ maps.  In 
the second step, we refine the individual local solutions by letting all parameters, including 
$s$ and $q$, vary. If the degeneracies among different local solutions are severe, we present
 multiple solutions, and otherwise we present a single global solution.

\subsection{KMT-2017-BLG-0673}\label{sec:three-one}

The anomaly feature of KMT-2017-BLG-0673, shown in Figure~\ref{fig:three}, is characterized by 
a single strong anomaly point at HJD$^\prime\sim 7963.9$ ($t_{\rm anom}$) surrounded by weak 
positive deviations around $t_{\rm anom}$ during the relatively short time range of $7960\lesssim 
{\rm HJD}^\prime \lesssim 7967$. The light curve variation at $t_{\rm anom}$ is discontinuous, 
and this suggests that the anomaly point was produced by the source crossing over a caustic, 
although the detailed structure of the caustic-crossing feature could not be delineated due to 
the sparse coverage of the anomaly caused by the low observational cadence of the field.

Considering that the anomaly appears in the peripheral part of the light curve, it is likely 
that the anomaly was generated by a planetary caustic lying at around the planet-host axis 
with a separation of $u_{\rm anom}\sim s-1/s$ from the host \citep{Griest1998, Han2006}.  In 
this case, the planet-host separation can be heuristically estimated as \begin{equation} s = 
{1\over 2} \left[(u_{\rm anom}^2 +4)^{1/2} \pm u_{\rm anom}\right], \label{eq1} \end{equation} 
where $u_{\rm anom}=(\tau_{\rm anom}^2+u_0^2)^{1/2}$, $\tau_{\rm anom}=(t_{\rm anom}-t_0)/\te$, 
and $t_{\rm anom}\sim 7964$ represents the time of the anomaly.  With the values of $t_0\sim 
7973.3$, $u_0\sim 0.12$, and $\te \sim 22$~days obtained from the 1L1S modeling conducted by 
excluding the data points around the anomaly, we find two values of $s_{\rm close}\sim 0.81$ 
and $s_{\rm wide}\sim 1.24$, which correspond to the separations of the close ($s<1.0$) and wide 
($s>1.0$) solutions, respectively.

From the 2L1S modeling, we found a unique solution without any degeneracy. In Table~\ref{table:one}, 
we list the lensing parameters of the solution and the model curve corresponding to the solution 
is drawn over the data points in Figure~\ref{fig:three}. It was found that the 2L1S model improved 
the fit by $\Delta\chi^2=584.7$ with respect to the 1L1S model. As expected, the anomaly was 
produced by a low-mass companion with the binary parameters of $(s, q)\sim (0.81, 5.6\times 
10^{-3})$. We note that the planet-host separation matches well the value that was heuristically 
estimated from the location of the anomaly in the lensing light curve.

\begin{table}[t]
\small
\caption{Lensing parameters of KMT-2017-BLG-0673\label{table:one}}
\begin{tabular*}{\columnwidth}{@{\extracolsep{\fill}}lllcc}
\hline\hline
\multicolumn{1}{c}{Parameter}    &
\multicolumn{1}{c}{Value }       \\
\hline
$\chi^2$                &   $452.0             $      \\
$t_0$ (HJD$^\prime$)    &   $7973.264 \pm 0.049$      \\
$u_0$                   &   $0.121 \pm  0.006  $      \\
$\te$ (days)            &   $22.36 \pm  0.63   $      \\
$s$                     &   $0.813 \pm  0.005  $      \\
$q$ ($10^{-3}$)         &   $5.58 \pm  1.22    $      \\
$\alpha$ (rad)          &   $3.179 \pm 0.032   $      \\
$\rho$ ($10^{-3}$)      &   $7.56 \pm  2.32    $      \\
\hline
\end{tabular*}
\tablefoot{ ${\rm HJD}^\prime = {\rm HJD}- 2450000$.  }
\end{table}

\begin{table*}[t]
\small
\caption{Lensing parameters of KMT-2019-BLG-0414\label{table:two}}
\begin{tabular}{llllll}
\hline\hline
\multicolumn{1}{c}{Parameter}   &
\multicolumn{2}{c}{2L1S (Sol 1)}  &
\multicolumn{2}{c}{2L1S (Sol 2)}  &
\multicolumn{1}{c}{Xallarap}    \\
\multicolumn{1}{c}{}            &
\multicolumn{1}{c}{Close}       & 
\multicolumn{1}{c}{Wide}        &
\multicolumn{1}{c}{Close}       &
\multicolumn{1}{c}{Wide}        &
\multicolumn{1}{c}{}        \\
\hline
$\chi^2$                &  $585.8             $   &  $585.8             $   & $585.9             $  &  $586.0             $  &  $590.0              $     \\
$t_0$ (HJD$^\prime$)    &  $8611.332 \pm 0.005$   &  $8611.330 \pm 0.005$   & $8611.333 \pm 0.005$  &  $8611.335 \pm 0.005$  &  $8611.362 \pm 0.015 $     \\
$u_0$ ($10^{-3}$)       &  $4.33 \pm 0.52     $   &  $4.31 \pm 0.49     $   & $4.48 \pm 0.53     $  &  $4.26 \pm 0.56     $  &  $4.84 \pm    0.63   $     \\
$\te$ (days)            &  $71.32 \pm 7.80    $   &  $71.31 \pm 7.70    $   & $70.77 \pm 7.50    $  &  $74.65 \pm 8.17    $  &  $72.28 \pm     4.48 $     \\
$s$                     &  $0.347 \pm 0.066   $   &  $2.803 \pm 0.599   $   & $0.416 \pm 0.078   $  &  $2.714 \pm 0.376   $  &  --                        \\
$\log q$                &  $-2.23 \pm 0.26    $   &  $-2.26 \pm 0.25    $   & $-2.60 \pm 0.28    $  &  $-2.56 \pm 0.26    $  &  --                        \\
$\alpha$ (rad)          &  $2.986 \pm 0.064   $   &  $3.003 \pm 0.056   $   & $0.783 \pm 0.067   $  &  $0.784 \pm 0.068   $  &  --                        \\
$\rho$ ($10^{-3}$)      &  $< 5               $   &  $< 5               $   & $< 5               $  &  $< 5               $  &  --                        \\
$P$ (day)               &                         &                         &                       &                        & $ 0.9 \pm 0.1        $     \\
$\xi_N$ ($10^{-3}$)     &                         &                         &                       &                        & $  0.84 \pm    0.16  $     \\
$\xi_E$ ($10^{-3}$)     &                         &                         &                       &                        & $  0.00 \pm    0.12  $     \\
$\phi$ (deg)            &                         &                         &                       &                        & $  326.72 \pm   15.39$     \\
$i$ (deg)               &                         &                         &                       &                        & $  11.83 \pm   12.24 $     \\
\hline                                                                                                                                                   
\end{tabular}
\end{table*}

The lens system configuration, showing the source motion with respect to the caustic, is 
presented in the inset of the top panel in Figure~\ref{fig:three}.  It shows that the anomaly 
was generated by the source passage through one of the two planetary caustics induced 
by a close planet with $s<1.0$. To be noted among the parameters is that the normalized 
source radius $\rho =(7.56\pm 2.32)\times 10^{-3}$ is measured despite the fact that 
only a single point covered the caustic. This is possible because the data points lying 
adjacent to the caustic point provide an extra constraint on the source radius.

A short-term anomaly can also be produced if a source is a binary \citep{Gaudi1998}.  We 
checked this possibility by additionally conducting a binary-source (1L2S) modeling of the 
light curve. From this modeling, it was found that the 1L2S model yielded a poorer fit to the 
anomaly than the 2L1S model by $\Delta\chi^2 =30.7$, especially in the peripheral region of 
the anomaly. We, therefore, exclude the 1L2S interpretation of the anomaly.  We also checked
the feasibility of detecting higher-order effects, including the microlens-parallax effect 
\citep{Gould1992} and lens-orbital effect \citep{Batista2011, Skowron2011} induced by the 
orbital motion of Earth and the binary lens, respectively.  We found that it was difficult 
to securely constrain the lensing parameters defining these higher-order effects not only 
because the event time scale was not long enough but also because the photometry was not 
sufficiently precise.

\subsection{KMT-2019-BLG-0414}\label{sec:three-two}

\begin{figure}[t]
\includegraphics[width=\columnwidth]{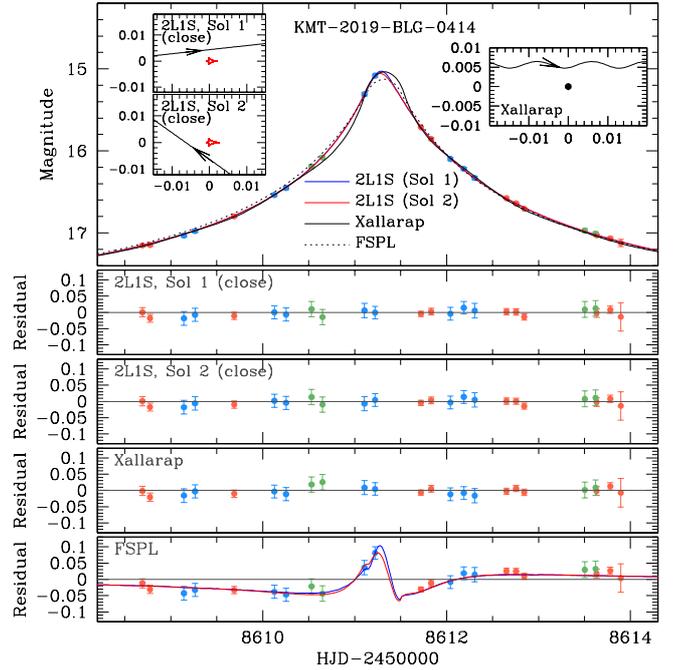}
\caption{
Enlarged view around the anomaly region in the lensing light curve of KMT-2019-BLG-0414. 
Drawn over the data points are the 1L1S model considering finite-source effects (FSPL), 
the two 2L1S (solutions~1 and 2) models, and   the binary-source model considering xallarap 
effects (xallarap).  The model curves of the two 2L1S models are difficult to be resolved 
within the line width due to the severe degeneracy between the models. The four bottom panels 
show the residuals from the individual models.  The two left insets in the top panel show 
the lens-system configurations of the two 2L1S solutions,  and the right inset shows the 
configuration of the xallarap model.  The blue and red curves in the bottom panel represent 
the differences of the 2L1S solutions 1 and 2 from the FSPL model, respectively.
}
\label{fig:four}
\end{figure}

The event KMT-2019-BLG-0414 reached a very high magnification at the peak: $A_{\rm peak} > 
200$.  The anomaly, whose features are shown in Figure~\ref{fig:four}, occurred near the peak, 
and the deviation from the 1L1S model was subtle without any strong signatures, such as would 
be generated by a caustic crossing. From the inspection of the residual from the 1L1S model, 
presented in the bottom panel, the feature of the anomaly is characterized by two KMTA points 
at the peak (HJD$^\prime =8611.11$ and 8611.22) with positive deviations and those around the 
peak exhibiting slight negative deviations that lasted for about 2.5 days.

\begin{table*}[t]
\small
\caption{Source properties\label{table:three}}
\begin{tabular}{lllll}
\hline\hline
\multicolumn{1}{c}{Quantity}             &
\multicolumn{1}{c}{KMT-2017-BLG-0673}    &
\multicolumn{1}{c}{KMT-2019-BLG-0414}    \\
\hline
$(V-I, I)_{\rm S}$        &  $(3.728 \pm 0.063, 17.291\pm 0.006)  $    &  $(2.603 \pm 0.029, 21.476 \pm 0.003)$      \\
$(V-I, I)_{\rm RGC}$      &  $(3.693, 17.517)                     $    &  $(2.966, 17.435)                    $      \\
$(V-I, I)_{\rm RGC,0}$    &  $(1.060, 14.377)                     $    &  $(1.060, 14.239)                    $      \\
$(V-I, I)_{\rm 0,S}$      &  $(1.096 \pm 0.063, 14.150 \pm 0.006) $    &  $(0.697 \pm 0.029, 18.280\pm 0.003) $      \\
$\theta_*$ ($\mu$as)      &  $7.29 \pm 0.70                       $    &  $0.68\pm 0.05                       $      \\
\hline
\end{tabular}
\end{table*}

From the 2L1S modeling of the light curve, we found 2 sets of planetary solutions with source 
trajectory angles of $\alpha \sim 3.00$ (in radian) and $\sim 0.78$.  We refer to the solutions 
as  2L1S ``solution~1'' and ``solution~2'', respectively. For each solution, we identified
a pair of solutions resulting from the close-wide degeneracy \citep{Griest1998, Dominik1999}, 
and thus there are 4 solutions in total. Regardless of the solution, the estimated mass ratios 
between the lens components are in the range of [3.3 -- 4.7]$\times 10^{-3}$, indicating that 
the companion to the lens is a planet. The fit improvement of the 2L1S models with respect to 
a 1L1S model conducted with the consideration of finite-source effects is $\Delta\chi^2= 90.8$, 
indicating that the planetary signal is securely detected.

In Table~\ref{table:two}, we list the lensing parameters of the 4 degenerate planetary 2L1S
solutions along with the $\chi^2$ values of the fits. The degeneracies among the solutions 
are very severe with $\Delta\chi^2 <0.2$. In Figure~\ref{fig:four}, we present the model 
curves and residuals of solutions~1 and 2. The lens system configurations of the solutions~1 
and 2 are shown in the left and right insets inserted in the top panel of Figure~\ref{fig:four}, 
respectively, where the presented configurations are for the close solutions. Despite the large 
difference in the source trajectory angle, $\alpha\sim 3.00$ for solutions~1 and $\sim 0.78$ for 
solution~2, the planet parameters, $(s, q)\sim (0.35/2.8, 4.7\times 10^{-3})$ for solution~1 and 
$(0.42/2.7, 3.3\times 10^{-3})$ for solution~2, are similar to each other, and thus the caustics 
of both solutions appear to be similar to each other. For both solutions, the anomaly was produced 
by the source approach close to the central caustic induced by a planetary companion. The two points 
with positive deviations appeared when the source approached the cusp of the caustic.  Although the 
normalized source radius cannot be measured because the source did not cross the caustic, the 
modeling yields an upper limit of $\rho_{\rm max}\sim  5 \times  10^{-3}$.

We additionally checked the binary-source origin the anomaly.  Under the static 1L2S 
interpretation without the orbital motion of the source, we found that the anomaly could 
not be explained by the model because the light curve exhibited both positive and negative 
deviations, while the 1L2S model could produce only positive deviations.  However, a negative 
deviation can be produced by a xallarap effect, in which a faint source companion induces 
a variation of the primary source motion by the orbital motion \citep{Griest1992, Han1997}.  
Following the parameterization of \citet{Dong2009}, we thus carried out a xallarap modeling 
by adding 5 extra parameters of ($\xi_{{\rm E},N}, \xi_{{\rm E},E}$, $P$, $\phi$, $i$) to 
those of the 1L1S model.  Here $(\xi_{{\rm E},N}, \xi_{{\rm E},E})$ are the north and east 
components of the xallarap vector $\xivec_{{\rm E}}$, $P$ denotes the orbital period, 
$\phi$ is the phase angle, and $i$ represents the inclination of the source orbit.  From 
this modeling, we found a xallarap solution that approximately explained the observed anomaly.  
We list the lensing parameters of the best-fit xallarap solution in Table~\ref{table:two}, 
and present the model curve, its residual, and the lens system configuration in 
Figure~\ref{fig:four}.  According to the model, the source is accompanied by a close faint 
companion inducing a very short orbital period of $\sim 1$~day.  Although the xallarap fit 
is not preferred over the the 2L1S solutions, the $\chi^2$ difference is minor with $\Delta
\chi^2\sim 4.2$.  We, therefore, consider the xallarap solution as a viable interpretation of 
the event.  Had the Einstein radius of the lens system been measured, the minimum mass of the 
source companion could be constrained by the relation of $M_{S_2,{\rm min}}=(\xi_{\rm E}
\hat{r}_{\rm E}/{\rm AU})^3/(P/{\rm yr})^2$ to judge the validity of the xallarap solution, 
but this constraint could not be applied to the lens system because the normalized source 
radius could not be measured.  Here $\hat{r}_{\rm E}=D_{\rm S}\thetae$ denotes the physical 
Einstein radius projected on the source plane.  We note that the model curves of the the 2L1S 
and xallarap solutions differ from each other in the region not covered by data points, 
indicating that the degeneracy between the solutions could have been resolved if the event 
had more continuous coverage. In particular, the two models differ by $\sim 0.1$~mag during 
the KMTC observing window centered on HJD$^\prime \sim 8610.75$. Unfortunately, KMTC was
weathered out on that night.

\section{Source stars and Einstein radii}\label{sec:four}

In this section, we characterize the source stars of the events.  Characterizing the source 
star is important not only to fully describe the event but also to measure the Einstein 
radius, which is related to the lensing parameter of the normalized source radius by
\begin{equation}
\thetae = {\theta_* \over \rho},
\label{eq2}
\end{equation}
where the angular source radius can be deduced from the source type. See below for the detailed 
procedure of the $\theta_*$ estimation. Measurement of the Einstein radius is important because it 
is related to the mass $M$ and distance to the lens $\dl$ as $\thetae=(\kappa M\pi_{\rm rel})^{1/2}$, 
and thus $\thetae$ can provide an extra constraint on the mass and distance to the lens in addition 
to the basic observable of the event time scale.  Here $\kappa\equiv 4 G/c^2\,{\rm AU}\simeq 
8.14\,{\rm mas}/M_\odot$ and $\pi_{\rm rel}$ represents the relative lens-source parallax, that is, 
$\pi_{\rm rel}={\rm AU}(D_{\rm L}^{-1} - D_{\rm S}^{-1})$, where $D_{\rm S}$ is the distance to the 
source.   We specified the source stars of the events by measuring their color and brightness.  In 
order to estimate the extinction- and reddening-corrected (de-reddened) color and magnitude of the 
source, $(V-I, I)_{\rm S,0}$, from the instrumental values, $(V-I, I)_{\rm S}$, we applied the 
\citet{Yoo2004} routine. In this routine, the centroid of the red giant clump (RGC) in the 
color-magnitude diagram (CMD) with its known de-reddened color and magnitude \citep{Bensby2013, 
Nataf2013} was used as a reference for the calibration of the source color and brightness.

\begin{figure}[t]
\includegraphics[width=\columnwidth]{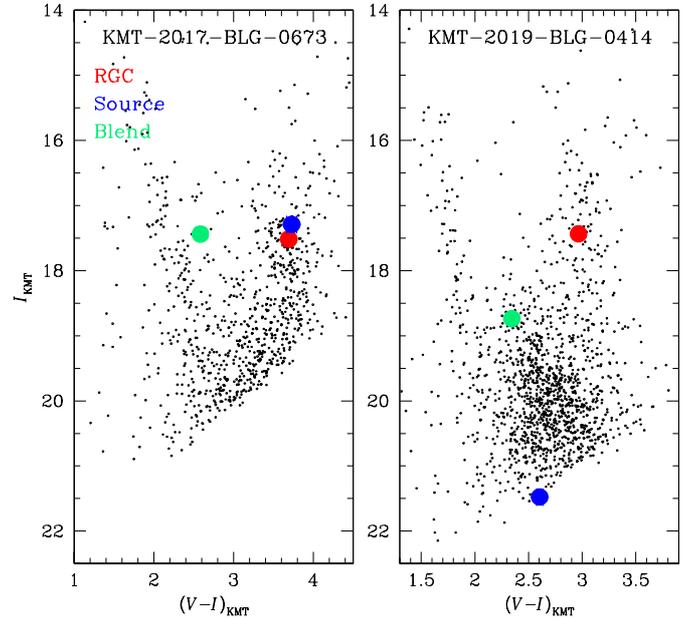}
\caption{
Source locations with respect to the centroids of red giant clump (RGC) in the instrumental 
color-magnitude diagrams of stars lying in the neighborhood of the source of KMT-2017-BLG-0673 
and KMT-2019-BLG-0414 constructed using the pyDIA photometry of KMTC data set. Also marked are 
the positions of the blend.
}
\label{fig:five}
\end{figure}

Figure~\ref{fig:five} shows the CMDs of stars lying in the neighborhood of the source stars 
of the two events constructed with the photometry data processed using the pyDIA code 
\citep{Albrow2017}. In each CMD, we mark the positions of the source and RGC centroid by a 
blue dot and a red dot, respectively.  Also marked is the position of a blend, which is marked 
by a green dot, indicating that the blended fluxes of both events are dominated by bright disk 
stars. We will discuss the nature the blend in the following section. For the estimation of the 
source position in the CMD, we measured the $I$- and $V$-band magnitudes of the source by regressing 
the data of the individual passbands processed using the pyDIA photometry code with the variation 
of the lensing magnification. By measuring the offsets in color and magnitude between the source 
and RGC centroid in the CMD, $\Delta (V-I, I) = [(V-I, I)_{\rm S} - (V-I, I)_{\rm RGC}]$, then 
the de-reddened color and magnitude of the source  were estimated as
\begin{equation}
(V-I, I)_{\rm S,0} = (V-I, I)_{\rm RGC,0} + \Delta (V-I, I). 
\label{eq3}
\end{equation}
In Table~\ref{table:three}, we list the values of $(V-I, I)_{\rm S}$, $(V-I, I)_{\rm RGC}$, 
$(V-I, I)_{\rm RGC,0}$, and $(V-I, I)_{\rm 0,S}$ for the two lensing events. According to the 
measured color and magnitude, it was found that the source of KMT-2017-BLG-0673 was a K-type 
giant and the source of KMT-2019-BLG-0414 was a G-type main-sequence star.

We estimated the source radii of the events by first converting $V-I$ into $V-K$ using the
color-color relation of \citet{Bessell1988}, and then derived 
$\theta_*$ from the 
$(V-K, V)$--$\theta_*$ relation of \citet{Kervella2004}.  The estimated angular radii of the 
source stars are
\begin{equation}
\theta_* =
\begin{cases}
(7.29 \pm 0.70)~\mu{\rm as}, & \textrm{for KMT-2017-BLG-0673}, \\
(0.68 \pm 0.05)~\mu{\rm as},  & \textrm{for KMT-2019-BLG-0414}.
\end{cases}
\label{eq4}
\end{equation}
From the relation in Equation~(\ref{eq2}), it is estimated that the angular Einstein radii of 
the two events are
\begin{equation}
\thetae =
\begin{cases}
(0.96 \pm 0.31)~{\rm mas}, & \textrm{for KMT-2017-BLG-0673}, \\
> 0.14~{\rm mas},          & \textrm{for KMT-2019-BLG-0414}.
\end{cases}
\label{eq5}
\end{equation}
With the values of the event time scales, the relative lens-source proper motions, 
$\mu=\thetae/\te$, are estimated as
\begin{equation}
\mu = 
\begin{cases}
(15.74 \pm 5.05)~{\rm mas/yr}, & \textrm{for KMT-2017-BLG-0673}, \\
> 0.7~{\rm mas/yr},            & \textrm{for KMT-2019-BLG-0414}.
\end{cases}
\label{eq6}
\end{equation}
For KMT-2019-BLG-0414, we present the lower limits of $\thetae$ and $\mu$ because only an 
upper limit on $\rho$ is obtained for this event.

\begin{table*}[t]
\small
\caption{Physical lens parameters\label{table:four}}
\begin{tabular}{lccccc}
\hline\hline
\multicolumn{1}{c}{Quantity}               &
\multicolumn{1}{c}{KMT-2017-BLG-0673}      &
\multicolumn{4}{c}{KMT-2019-BLG-0414}      \\
\multicolumn{1}{c}{}                       &
\multicolumn{1}{c}{}                       &
\multicolumn{2}{c}{Solution 1}             &
\multicolumn{2}{c}{Solution 2}             \\
\multicolumn{1}{c}{}                       &
\multicolumn{1}{c}{}                       &
\multicolumn{1}{c}{Close}                  &
\multicolumn{1}{c}{Wide}                   &
\multicolumn{1}{c}{Close}                  &
\multicolumn{1}{c}{Wide}                   \\
\hline
$M_{\rm h}$ ($M_\odot$)      &  $0.63^{+0.37}_{-0.35}$   &  $0.74^{+0.43}_{-0.38}$   & $\leftarrow          $  &  $\leftarrow          $  &  $\leftarrow           $ \\  [0.7ex]
$M_{\rm p}$ ($M_{\rm J}$)    &  $3.67^{+2.17}_{-2.07}$   &  $4.57^{+3.74}_{-2.06}$   & $4.26^{+3.32}_{-1.87}$  &  $1.95^{+1.76}_{-0.93}$  &  $2.14^{+1.75}_{-0.96} $ \\  [0.7ex]
$\dl$ (kpc)                  &  $5.08^{+1.22}_{-1.59}$   &  $4.41^{+1.79}_{-1.93}$   & $\leftarrow          $  &  $\leftarrow          $  &  $\leftarrow           $ \\  [0.7ex]
$a_\perp$ (AU)               &  $2.34^{+0.56}_{-0.74}$   &  $1.16^{+0.47}_{-0.51}$   & $9.42^{+3.83}_{-4.12}$  &  $1.40^{+0.57}_{-0.61}$  &  $9.12^{+3.71}_{-3.99} $ \\  [0.7ex]
disk/bulge                   &  $63\%/37\%           $   &  $73\%/27\%           $   & $\leftarrow          $  &  $\leftarrow          $  &  $\leftarrow           $ \\  [0.7ex]
\hline
\end{tabular}
\end{table*}

\begin{figure}[t]
\includegraphics[width=\columnwidth]{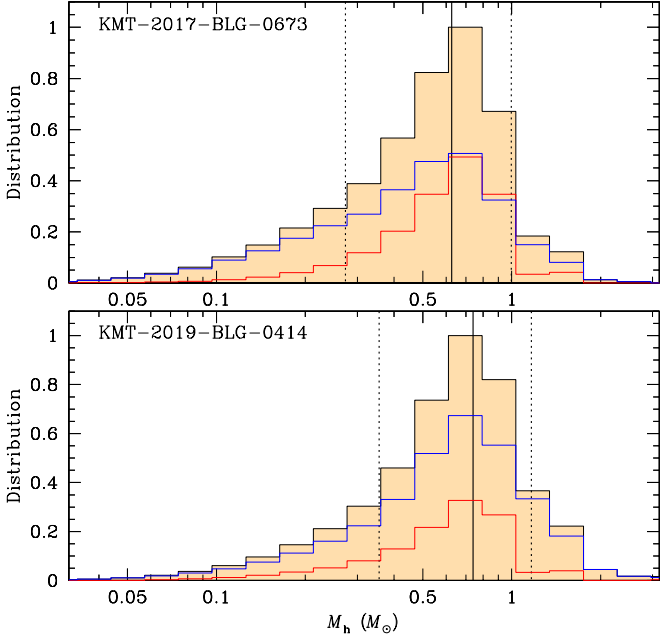}
\caption{
Bayesian posteriors of the host mass for the planetary systems KMT-2017-BLG-0673L (upper panel) and 
KMT-2019-BLG-0414L (lower panel). The vertical line in each panel represents the median value and the 
two dotted vertical lines indicate the 1$\sigma$ range of the distribution. The curves drawn in blue 
and red represent the contributions of the disk and bulge lens populations, respectively.
}
\label{fig:six}
\end{figure}

\section{Physical lens parameters}\label{sec:five}

The physical parameters of the lens mass and distance are constrained by the lensing observables
of $\te$, $\thetae$, and $\pie$, which are related to the physical parameters.  The basic parameter 
of the event time scale was securely measured for both events, the Einstein radius was measured for 
KMT-2017-BLG-0673, while only a lower limit was obtained for KMT-2019-BLG-0414.  However, the microlens 
parallax could not be measured for either event.  Although the partial measurements of the lensing 
observables make it difficult to uniquely determine $M$ and $\dl$ from the relations \citep{Gould2000}
\begin{equation}
M= {\thetae \over \kappa\pie};\qquad 
\dl ={{\rm AU} \over \pie\thetae + \pi_{\rm S}},
\label{eq7}
\end{equation}
it is still possible to constrain the physical parameters with the measured observables of the
individual events from a Bayesian analysis.

\begin{figure}[t]
\includegraphics[width=\columnwidth]{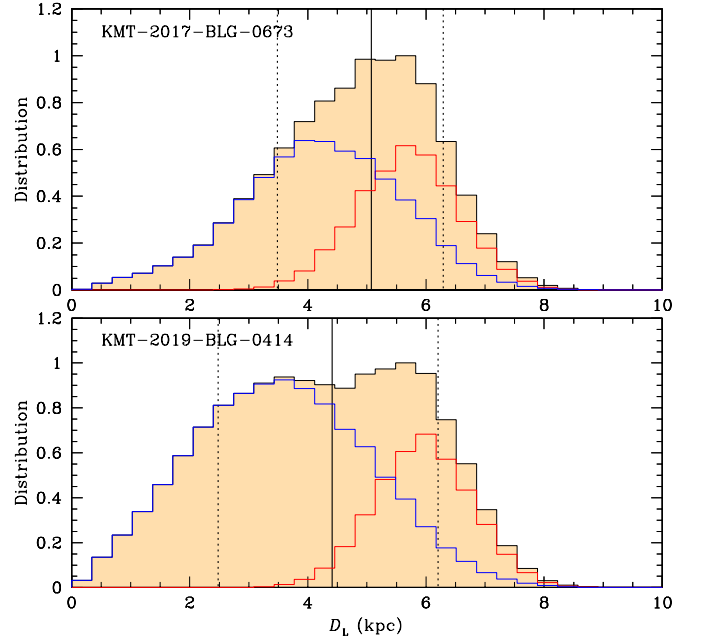}
\caption{
Bayesian posteriors of the distance to the planetary system. Notations are same as those in
Fig.~\ref{fig:six}.
}
\label{fig:seven}
\end{figure}

In the Bayesian analysis, we first produced a large number ($10^7$) of lensing events by conducting 
a Monte Carlo simulation based on a Galactic model. The Galactic model defines the matter density,
motion, and mass function of Galactic objects. From the physical parameters of the simulated events, 
we computed the lensing observables of $t_{{\rm E},i}=\dl \thetae/v_\perp$ and $\theta_{{\rm E},i}
=(\kappa M \pi_{\rm rel})^{1/2}$.  Here $v_\perp$ represents the transverse speed between the lens 
and source.  We then constructed Bayesian posteriors of the lens mass and distance 
with a weight assigned to each simulated event computed by $w_i=\exp(-\chi^2/2)$, where $\chi^2=
(t_{{\rm E},i}-\te)^2/\sigma({\te})^2+(\theta_{{\rm E},i}-\thetae)^2/\sigma(\thetae)^2$, $\sigma(\te)$ 
and $\sigma(\thetae)$ represent the uncertainties of $\te$ and $\thetae$ estimated from the modeling, 
respectively. In the simulation of lensing events, we used the \citet{Jung2021} Galactic model.

In Figures~\ref{fig:six} and \ref{fig:seven}, we present the posteriors of the host mass and distance 
to the planetary systems, respectively. For each distribution, the solid vertical line denotes the 
median value, and the two dotted vertical lines represent the 1$\sigma$ range estimated as 16\% and 
84\% of the distribution. We separately mark posterior distributions of the disk and bulge lens 
populations with blue and red curves, respectively. In Table~\ref{table:four}, we summarize the 
estimated masses of the host ($M_{\rm h}$) and planet ($M_{\rm p}$), distances, and projected 
planet-host separations ($a_\perp=s\dl\thetae$) for the two planetary systems. For KMT-2019-BLG-0414, 
we present four sets of parameters corresponding to the four degenerate solutions. We note that the 
primary mass and distance of KMT-2019-BLG-0414L estimated from the degenerate solutions are alike 
regardless of the solution because the constraint given by the lower limit of $\thetae$ is very weak 
and the other constraint of the event time scales of the degenerate solutions are nearly identical.

It is found that the two planetary systems are very similar to each other in many aspects. First, 
both systems have planets heavier than the Jupiter of the Solar system with masses $M_{\rm p}\sim 
3.7~M_{\rm J}$ and $\sim [2.0$--$4.5]~M_{\rm J}$ for KMT-2017-BLG-0673L and KMT-2019-BLG-0414L, 
respectively. Second, the masses of the planet hosts, $M_{\rm h}\sim 0.6~M_\odot$ and $\sim 
0.7~M_\odot$ for the individual lenses, are also similar to each other. Third, the distances to 
the systems, $\dl\sim 5.1$~kpc and $\sim 4.4$~kpc, indicate that both planetary systems are likely 
to be in the disk. According to the Bayesian estimation, the probability for the planetary system 
to be in the disk is $\sim 63\%$ for KMT-2017-BLG-0673L and $\sim 73\%$ for
KMT-2019-BLG-0414L.

Prompted by the facts that the lenses of both events are likely to be in the disk and the blended
fluxes come from disk stars, we inspected the possibility of the lenses of the events to be the 
main sources of the blended flux. For this inspection, we examined the reference images taken 
before the lensing magnification and the difference image taken during the lensing magnification. 
For KMT-2017-BLG-0673, there exists a star even brighter than the source of a bulge giant at a
location with a separation of $\sim 2.2$~arcsec from the source, and this degrades the photometry 
and astrometry of the event together with the high extinction, $A_I\sim 3.1$~mag, toward the field. 
Nevertheless, we identified a blend lying at a separation of $\sim 0.65$~arcsec from the source with 
a brightness corresponding to the blend. This indicates that the lens is not the main source of the 
blended flux. For KMT-2019-BLG-0414, the astrometric offset between the centroid of baseline object 
measured on the reference image, $\theta_{\rm ref}=[(f_s/f_b)\theta_s+\theta_b]/(f_s/f_b+1)$, and 
the source position measured on the difference image, $\theta_s$, is $\Delta\theta=|\theta_{\rm ref}
-\theta_s|=(79.2\pm 59.2)$~mas.  Here $f_s$ and $f_b$ represent the flux values of the source and 
blend, respectively.  In the case of KMT-2019-BLG-0414, $f_b\gg f_s$, and thus the centroid shift 
can be approximated as $\Delta\theta\sim |\theta_b-\theta_s|$.  Considering that the measured 
astrometric offset is only 1.3 times greater than the uncertainty, it is difficult to firmly 
exclude the possibility that the lens is the major source of blended light based on the argument 
of the measurement uncertainty.

\section{Resolution of KMT-2019-BLG-0414 Degeneracy}\label{sec:six}

In this section, we discuss the possibility of resolving the degeneracy between the 2L1S and 
xallarap solutions of KMT-2019-BLG-0414 by means of AO observations on 30~m class telescopes, 
which may see first light in about 2030.  We give a brief synopsis of the decision tree governing 
these observations.

In the 2L1S model, $\rho<0.005$ at $3\,\sigma$ confidence, which implies, $\mu>0.7$~mas~yr$^{-1}$
(see Equation~(\ref{eq6})). If $\mu$ is really at the threshold of this limit, then several
decades would be required to separately resolve the lens and source. However, typical lens-source
relative proper motions are substantially higher than this, and therefore, most likely, 30~m-class 
AO imaging in 2030 will resolve them. This will yield a heliocentric proper motion measurement. 
One must be careful to correct from heliocentric to geocentric proper motion using
\begin{equation}
\muvec_{\rm geo} = \muvec_{\rm hel} - {{\bf v}_{\oplus,\perp}\over {\rm AU}}\pi_{\rm rel},
\label{eq8}
\end{equation}
where ${\bf v}_{\oplus,\perp}(N,E) = (-0.7,+21.6)~{\rm km~s}^{-1}$ is the projected velocity 
of Earth at the peak of the event. However, we ignore this correction in the following
simplified treatment. Because $\te \simeq {\rm yr}/5$ is well measured in both models, such 
a proper motion measurement will immediately yield a determination of the Einstein radius: 
$\thetae \simeq 0.6~{\rm mas}~(\mu/3\,{\rm mas}~{\rm yr}^{-1})$.

Within the context of the xallarap model, this will yield the semi-major axis of the orbit of 
the primary, i.e., $a_{\rm primary}=\xi\theta_\e D_{\rm S}=1.25~R_\odot~(\mu/3~{\rm mas~yr}^{-1})$, 
where we have assumed a source distance of $D_{\rm S}=8$~kpc. Combined with the period measurement 
$P=0.9$~day, a source-mass estimate of $M_{\rm S} \simeq 1\,M_\odot$ and Kepler's Third Law, this 
will yield the mass of the source companion. For example, for $\mu=3~{\rm mas~yr}^{-1}$, 
$M_{\rm companion}\sim 0.4\,M_\odot$ (which would account for its lack of photometric signature). 
However, if $\mu$ were measured to be sufficiently large, this might imply photometric signatures 
from a luminous orbiting companion and (or) eclipses, which might rule out the xallarap scenario.

Assuming that xallarap solutions survive this test, there remains a radial-velocity (RV) test 
of the xallarap hypothesis. The amplitude of the RV signature would be $v\sin i = 2\pi (a_{\rm 
primary}/P) \sin i = 70~{\rm km~s}^{-1} (\mu/3~{\rm mas~yr}^{-1})\sin i$.  The source is 
$I_{\rm S} \sim 21.5$ ($K_{\rm S} \sim 18.0$), so RV signatures of amplitude $\ll 1~{\rm km~s}^{-1}$ 
should be detectable with 30~m-class AO spectroscopy. If these signatures are not detected, then 
the xallarap model would become highly implausible because it would require an almost perfectly 
face-on orbit.  For completeness, we note that {\it Gaia} DR3 reports a parallax and proper 
motion for the baseline object of $\pi_{\rm base} = (1.1\pm 0.4)$~mas and $\muvec_{\rm base}(E,N) 
= (-0.4\pm 0.4,-2.3\pm 0.3)~{\rm mas~yr}^{-1}$. As shown in Figure~\ref{fig:five}, the baseline 
object is dominated by the blend.  While the parallax error is large, the low proper motion is 
also consistent with the blend being a nearby object. This object is most likely dominated by 
the host or a companion to the host. By the time 30~m-class AO is available, the parallax error 
will likely be reduced by a factor $\sim 1.7$, so Gaia astrometry may substantially aid the 
interpretation of the AO images.

\section{Summary}\label{sec:seven}

We presented analyses of two microlensing events KMT-2017-BLG-0673 and KMT-2019-BLG-0414, for 
which the presence of planets in the lenses were found from the inspection of the microlensing 
data collected by the KMTNet survey during the 2017--2019 seasons in the peripheral Galactic
bulge fields.  The planetary signal of KMT-2017-BLG-0673 had been previously missed because of 
the relatively sparse coverage of the event, and the signal of KMT-2019-BLG-0414 had not been 
identified due to the weakness of the signal caused by its non-caustic-crossing nature.  The 
detections of the planets indicate the need for thorough investigations of non-prime-field 
lensing events for the complete census of microlensing planet samples.

For KMT-2017-BLG-0673, we identified a unique planetary solution with lensing parameter of $(s, q) 
\sim  (0.81, 5.6\times 10^{-3})$.  It was found that the planetary signal was generated by the 
source passage through one of the two tiny planetary caustics induced by a planet lying inside 
the Einstein ring of the planetary system.

For KMT-2019-BLG-0414, on the other hand, we identified two pairs of solutions, for each of which 
there were two solutions caused by the close-wide degeneracy.  Despite the fact that the incidence 
angles of the two sets of solutions differed from each other, the planet parameters of the projected 
planet-host separations and mass ratios were similar to each other.  Although slightly less favored, 
the anomaly could also be explained by a model, in which a single lens mass magnified a rapidly 
orbiting binary source with a faint companion. In Section~\ref{sec:six}, we discussed how the 
planet/xallarap degeneracy might one day be resolved by AO observations on 30~m class telescopes. 
Pending such resolution, KMT-2019-BLG-0414Lb should not be included in catalogs of 
``known planets.''

From the physical parameters estimated by conducting Bayesian analyses based on the observables of 
the events, it was found that the two planetary systems were similar to each other in many aspects.  
For both planetary systems, the hosts of the planets are stars with masses lower than the Sun, and 
they host planets heavier than the Jupiter of the Solar system.  Furthermore, both planetary systems 
reside in the disk of the Galaxy.

\begin{acknowledgements}
Work by C.H. was supported by the grants of National Research Foundation of Korea 
(2020R1A4A2002885 and 2019R1A2C2085965).
This work was financially supported by the Research Year of Chungbuk National University in 2021.
This research has made use of the KMTNet system operated by the Korea Astronomy and Space 
Science Institute (KASI) and the data were obtained at three host sites of CTIO in Chile, 
SAAO in South Africa, and SSO in Australia.
J.C.Y.  acknowledges support from NSF Grant No. AST-2108414.
W.Z. and H. Y. acknowledge the support by the National Science Foundation of China (Grant
No. 12133005).
\end{acknowledgements}

\end{document}